\documentclass[12pt,reqno]{amsart}

\usepackage[T1]{fontenc}
\usepackage{lmodern}
\usepackage{amsmath,amssymb,bm}
\usepackage{graphicx}
\usepackage{microtype}
\usepackage{cite}
\usepackage{xcolor}
\usepackage[colorlinks=true,linkcolor=blue,citecolor=blue,urlcolor=blue]{hyperref}

\numberwithin{equation}{section}
\allowdisplaybreaks

\newcommand{\dd}{\mathrm{d}}
\newcommand{\e}{\mathrm{e}}

\hypersetup{
  pdftitle={Kasner critical scaling in Einstein--Maxwell--scalar black holes with nonlinear collapse and scalarization},
  pdfauthor={Chihang He and Chao Liu}
}

\begin{document}

\title[Kasner critical scaling in EMS black holes]{Kasner critical scaling in Einstein--Maxwell--scalar black holes with nonlinear collapse and scalarization}

\author{Chihang He}
\address{School of Mathematics and Statistics,
Huazhong University of Science and Technology,
Wuhan 430074, Hubei Province, China}

\email{hechihang@foxmail.com}

\author{Chao Liu}
\address{School of Mathematics and Statistics,
Huazhong University of Science and Technology,
Wuhan 430074, Hubei Province, China}
\address{Center for Mathematical Sciences,
Huazhong University of Science and Technology,
Wuhan 430074, Hubei Province, China}

\email{chao.liu.math@foxmail.com}

\date{}

\begin{abstract} 

    A critical scaling exponent of $1/2$ was numerically observed
for the Kasner parameter of asymptotically flat scalarized
black holes and conjectured to be universal
[Phys.\ Rev.\ Lett.\ \textbf{136}, 251402 (2026)].
In the same static, spherically symmetric
Einstein--Maxwell--scalar setting, we show that this exponent
is not universal.
We develop a rigorous mathematical framework and construct
couplings realizing exponents $1/(2m)$ for any integer $m\ge1$
by tuning higher-order terms at fixed quadratic coupling
parameter $\mu$.
Nonlinear interior collapse produces a spacelike Kasner
singularity with $\beta_{\rm K}\propto\varepsilon^{-1}$,
where $\varepsilon$ is the value of scalar field at the event horizon. An exterior branch satisfying
$q/q_\star-1\propto\varepsilon^{2m}$,  
therefore, yields
$\beta_{\rm K}\propto|q/q_\star-1|^{-1/(2m)}$,
where $q$ is the charge-to-mass ratio and $q_\star$ its
critical value.
We also determine the prefactor explicitly.
This recovers the $1/2$ law and establishes a hierarchy
including $1/4$ and $1/6$.

\end{abstract}

\maketitle

\section{Introduction}
In recent years, black holes with nonlinear scalar couplings have attracted considerable interest \cite{HerdeiroEtAl2018ChargedScalarization,FernandesEtAl2019CouplingDependence,AstefaneseiEtAl2019EMSClasses,Hod2020ExistenceLine,DonevaEtAl2024ScalarizationReview,BelkhadriaPombo2024MixedScalarization}. Scalar fields can change the causal structure inside black holes and the geometry near their singularities \cite{HartnollEtAl2021Diving,DiasHorowitzSantos2021InsideAF,
Henneaux2022FinalKasner}. Mass inflation near the inner Cauchy horizon illustrates the sensitivity of charged black-hole interiors to perturbations
\cite{PoissonIsrael1990InternalStructure,Ori1991MassInflation}. In specific models, static black holes with scalar hair have no smooth inner Cauchy horizon \cite{CaiLiYang2021NoInnerHorizon,DiasHorowitzSantos2021InsideAF}. Detailed interior studies have found a rapid collapse of the Einstein--Rosen bridge followed by an approach to a spacelike Kasner singularity
\cite{HartnollEtAl2021Diving,DiasHorowitzSantos2021InsideAF,
VanDeMoortel2024ViolentCollapse}.
Near such a singularity, spatial scale factors follow
asymptotic power laws in the remaining proper time, with
Kasner exponents characterizing the anisotropic geometry
\cite{Henneaux2022FinalKasner}.
This raises a quantitative question: how does the geometry
near the singularity depend on the scalar coupling and on
exterior black-hole parameters?

For static, spherically symmetric, asymptotically flat
scalarized black holes in Einstein--Maxwell--scalar theory,
Li, Sun, and Yang 
\cite{LiSunYang2025CriticalInterior} numerically found the critical scaling law
$\beta_K\propto|q/q_\star-1|^{-1/2}$. They further conjectured that this exponent is universal
across scalarized models with a leading quadratic scalar
coupling.
Here $\beta_K$ determines the Kasner exponents, $q=Q/M$ is the
charge-to-mass ratio, and $q_\star$ is its value where the
scalarized branch bifurcates from a
Reissner--Nordstr\"om (RN) solution. Here we show analytically how higher-order terms in the
scalar--electromagnetic coupling function select the critical
exponent through the nonlinear exterior bifurcation.
Within a suitable class of even coupling functions, successive
fine-tuning yields exponents $1/(2m)$ for any integer $m\geq1$,
without changing  the quadratic coupling parameter $\mu$. We answer this conjecture in the negative by rigorously
constructing couplings that realize the $1/(2m)$ scaling laws.

\begin{figure}[!htbp]
 \centering
 \includegraphics[width=\textwidth]{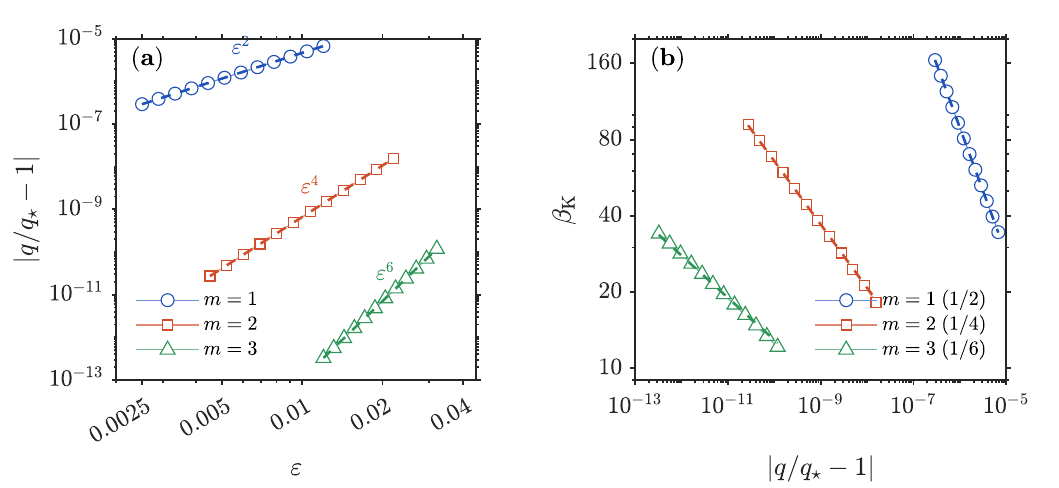}
\caption{Exterior contact and Kasner scaling for the couplings
in \eqref{eq:couplings} at $\mu=1$.\\
(a) $|q/q_\star-1|$ versus the horizon scalar value
$\varepsilon$; Eq.  \eqref{eq:factorization} gives exponents
$2$, $4$, and $6$ for $m=1,2,3$, respectively.\\
(b) $\beta_{\rm K}$ versus $|q/q_\star-1|$;
 Eq. \eqref{eq:intro_critical_scaling} gives exponents
$-1/2$, $-1/4$, and $-1/6$ for $m=1,2,3$, respectively.}
\label{fig:exponents}
\end{figure}

This hierarchy follows from two asymptotic relations in the
	horizon scalar value $\varepsilon>0$.
	Under the coupling assumptions stated below, we obtain
	\begin{equation}
		 \Delta q \equiv \frac{q(\varepsilon)}{q_\star}-1
 \sim c_{2m}\varepsilon^{2m},
 \qquad c_{2m}\ne0.\label{eq:factorization}
    \end{equation}
\begin{equation}
	 \beta_K \sim \frac{\lambda_I}{\iota}\varepsilon^{-1} =\frac{\pi}{\cosh\!\left(\pi\sqrt{\mu-\tfrac14}\right)} \varepsilon^{-1} ,
		\label{eq:intro_amplitude_scalings}
 \end{equation}
	as $\varepsilon\to0^+$.  
	Eliminating $\varepsilon$ gives
	\begin{equation}
		\beta_K \sim
		\frac{\lambda_I}{\iota}
		|c_{2m}|^{1/(2m)}
		|\Delta q|^{-1/(2m)},
		\label{eq:intro_critical_scaling}
	\end{equation} 
    where $\lambda_I/\iota=\pi/\cosh\!\left(\pi\sqrt{\mu-\tfrac14}\right)$
and $2m$ denotes the contact order. Higher-order coupling terms therefore select the critical
	exponent by changing the contact order, while
	leaving the leading interior law unchanged. We establish this hierarchy analytically and provide numerical
evidence confirming the predicted scaling laws.
	Fig.~\ref{fig:exponents} illustrates quadratic, quartic, and sextic contact,
	corresponding to exponents $1/2$, $1/4$, and $1/6$, respectively. 

\section{Model and observables}
We study static,
spherically symmetric, asymptotically flat black holes with
a purely electric field in four-dimensional
Einstein--Maxwell--scalar theory, with action
\begin{equation} 
S =   \frac{1}{16\pi}\int d^4x\,\sqrt{-g}\,
\bigl[R-2(\nabla\psi)^2  -Z(\psi)F_{\mu\nu}F^{\mu\nu}\bigr],
\quad F=dA . 
\label{eq:action}
\end{equation}
Here $\psi$ is a real scalar field, $A$ is the electromagnetic
potential, and $Z(\psi)>0$ is the scalar--electromagnetic
coupling function.
We impose $Z(0)=1$ and $Z_{,\psi}(0)=0$, so that the theory
admits the RN family with $\psi=0$.
For suitable choices of $Z$, branches of scalarized black
holes bifurcate from this family
\cite{DonevaEtAl2024ScalarizationReview,
HerdeiroEtAl2018ChargedScalarization}.  In this Letter, we let $W=Z^{-1}$ be even and
$C^{2m+4}$ near $\psi=0$, with $W(0)=1$, and set
$\mu=-W_{,\psi\psi}(0)/2>1/4$. We point out that
$\mu >1/4$
is necessary and sufficient for the Kasner critical scaling derived
here. A complete proof, which requires  technically intricate mathematical analysis,
is given in  \cite{CompanionMath}.

In event-horizon-radius units we write
\begin{equation}
 \begin{aligned}
 \dd s^2&=z^{-2}\left[-f\e^{-2\chi}\dd t^2
 +\frac{\dd z^2}{f}+\dd\Omega_2^2\right],\\
 \psi&=\psi(z),\qquad A=A_t(z)\dd t ,
 \end{aligned}
 \label{eq:ansatz}
\end{equation}
where $z=r_H/r$.  Spatial infinity is at $z=0$ and the event horizon at $z=1$.  We impose $\psi(0)=0$ and $f(1)=0$, and fix the time normalization by $\chi(1)=0$.  The signed branch parameter is $\varepsilon=\psi(1)$.  The quantities $Q$ and $M$ are the charge and ADM mass
in event-horizon-radius units, so $q=Q/M=Q_{\rm phys}/M_{\rm ADM}$. Let $Q_\star\in(0,1)$ denote the critical charge at which
the scalarized branch bifurcates from the RN solution.
The corresponding mass and charge-to-mass ratio are
$M_\star=(1+Q_\star^2)/2$ and
$q_\star=Q_\star/M_\star=2Q_\star/(1+Q_\star^2)$.

Inside the event horizon set $x=\ln z$ and $ \bar R=\e^{-\bar\chi}$, $ \bar N=-\e^{-3x}\bar R\bar f$ and $\bar J=\bar N\,\partial_x\bar\psi$. 
For each fixed sufficiently small $\varepsilon>0$,
we derive finite positive limits $ \bar N_c=\lim_{x\to\infty}\bar N(x;\varepsilon)$ and $ \bar J_c=\lim_{x\to\infty}\bar J(x;\varepsilon)$. Then we define the Kasner parameter by
$\beta_{\rm K}=\bar J_c/\bar N_c$.
As $x\to\infty$, the scalar field satisfies
$\bar\psi(x)\sim\beta_{\rm K}x$.
Let $\tau$ denote the remaining proper time to the singularity.
As $\tau\to0^+$, the metric and scalar field have the
asymptotic form
\begin{align}
 \dd s^2&\sim-\dd\tau^2+C_t\tau^{2p_t}\dd t^2
 +C_\Omega\tau^{2p_\Omega}\dd\Omega_2^2,\nonumber\\
 \bar\psi&\sim-\kappa_\psi\ln\tau,
 \label{eq:kasner-metric}
\end{align}
with
\begin{equation}
 p_t=\frac{\beta_{\rm K}^2-1}{\beta_{\rm K}^2+3},\quad
 p_\Omega=\frac{2}{\beta_{\rm K}^2+3},\quad
 \kappa_\psi=\frac{2\beta_{\rm K}}{\beta_{\rm K}^2+3}.
 \label{eq:kasner-exponents}
\end{equation}
Thus $\beta_{\rm K}$ determines the terminal geometry.

\section{Exterior contact}
For fixed $\mu>1/4$, we consider even couplings satisfying
$Z(\psi)=1+\mu\psi^2+O(\psi^4)$ near $\psi=0$.
We show that the higher terms in $Z$ can be chosen to obtain
contact order $2m$ for any prescribed integer $m\ge1$.

\begin{figure}[!htbp]
 \centering
 \includegraphics[width=\textwidth]{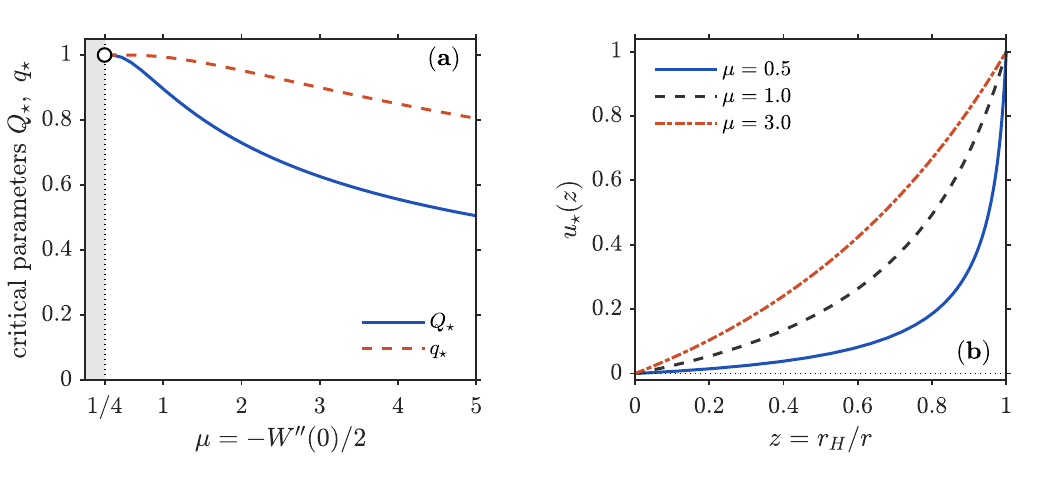}
 \caption{RN scalarization threshold and node-free linearized
scalar solutions.\\
(a) Scalarization occurs at a critical charge
$Q_\star\in(0,1)$ if and only if $\mu>1/4$.
The critical charge $Q_\star$ and charge-to-mass ratio
$q_\star$ decrease as $\mu$ increases.
Both approach $1$ as $\mu\to(1/4)^+$.\\
(b) For $\mu=0.5,1.0,3.0$, the normalized node-free solutions
satisfy $u_\star(0)=0$, $u_\star(1)=1$, and remain positive
on $0<z\le1$.}
\label{fig:threshold}
\end{figure}

Firstly, we point out the quadratic coupling parameter
$\mu$ in $Z$ determines the critical RN data $Q_\star$. Substituting $\psi(z)=\varepsilon u(z)$ and the RN values into
\eqref{eqA:exterior}, and retaining terms linear in
$\varepsilon$, gives
\begin{equation}
 \begin{gathered}
  \frac{\dd}{\dd z}
  \left[g_Q(z)\frac{\dd u(z)}{\dd z}\right]
  +\mu Q^2u(z)=0,\\
  g_Q(z)=(1-z)(1-Q^2z).
 \end{gathered}
 \label{eq:zero-mode}
\end{equation}
We denote by $u_\star$ the node-free exterior solution
at $Q=Q_\star$, satisfying $u_\star(0)=0$,
$u_\star(1)=1$, and $u_\star(z)>0$ for $0<z\le1$.
For each fixed $\mu>1/4$, we can prove that this equation
and these conditions uniquely determine $Q_\star$ and
$u_\star$, and hence $q_\star=2Q_\star/(1+Q_\star^2)$
(see \cite[\S3]{CompanionMath} for details). Fig.~\ref{fig:threshold} shows their dependence on $\mu$
and the node-free solution $u_\star$ for $\mu=0.5,1.0,3.0$. 
These quantities are therefore independent of the
higher terms in $Z$.

Next, let us show how these higher terms of $Z$ affect $q(\varepsilon)$.
It is convenient to expand $W=Z^{-1}$ at $\psi=0$:
\begin{equation}
 W(\psi)=\frac{1}{Z(\psi)}
 =1-\mu\psi^2+\sum_{k=2}^{m+1}\omega_{2k}\psi^{2k}
 +O(\psi^{2m+4}).
 \label{eq:coupling-expansion}
\end{equation}
For the scalarized branch, the symmetry $\psi\mapsto-\psi$
implies $q(-\varepsilon)=q(\varepsilon)$. This gives the finite
expansion as $\varepsilon\to0$, 
$ \frac{q(\varepsilon)}{q_\star}-1
 =\sum_{n=1}^{m}c_{2n}\varepsilon^{2n}
 +O(\varepsilon^{2m+2})$.
By Appendix \ref{app:tuning}, the dependence of $c_{2n}$ on the coupling is given by  
\begin{equation}
 c_{2n}=\Gamma_n+\varkappa_n\omega_{2n+2},
 \qquad \varkappa_n>0.
 \label{eq:triangular}
\end{equation}
Here $\Gamma_n$ depends on lower-order coupling and branch
coefficients, whereas $\varkappa_n$ depends only on $n$
and $\mu$ (see \cite{CompanionMath} for the lengthy expressions of $\Gamma_n$ and $\varkappa_n$).
Since $\varkappa_n>0$, choosing
$\omega_{2n+2}=-\Gamma_n/\varkappa_n$ sets $c_{2n}=0$
without changing the lower-order coefficients.
We make these choices successively for $n=1,\ldots,m-1$
and then choose $\omega_{2m+2}\ne-\Gamma_m/\varkappa_m$ so that $c_{2m}\ne0$.
Thus $c_2=\cdots=c_{2m-2}=0$, while $\mu$ and $u_\star$
remain fixed. The resulting branch satisfies \eqref{eq:factorization} as $\varepsilon\to0^+$. 

To verify the above theory, we provide the following three numerical examples for $\mu=1$ with contact orders $2$, $4$, and $6$, corresponding to $m=1,2,3$:
\begin{gather}
 Z_1 =1+\psi^2,\quad
 Z_2 =1+\psi^2+\gamma_\star\psi^4,
 \label{eq:couplings}\\
 Z_3 =1+\psi^2+\gamma_\star\psi^4+\delta_\star\psi^6.
 \nonumber
\end{gather}
For $Z=1+\psi^2+\gamma\psi^4+\delta\psi^6$, expanding
$W=Z^{-1}$ gives
\begin{equation}
 \omega_4=1-\gamma,\qquad
 \omega_6=-1+2\gamma-\delta.
 \label{eq:omega-gamma-delta}
\end{equation}
Thus $\gamma$ controls $\omega_4$. Once $\gamma$ is fixed,
varying $\delta$ changes $\omega_6$ without changing
$\omega_4$.

\subsection*{Step 1. Determine \texorpdfstring{$Q_\star$}{Q star} and \texorpdfstring{$u_\star$}{u star}.}

We determine $Q_\star$ by a numerical shooting method.
For each $Q\in(0,1)$, we integrate
\eqref{eq:zero-mode} from $z=0$ with $u(0)=0$ and
$\left.\dd u/\dd z\right|_{z=0}=1$ toward $z=1$.
We adjust $Q$ to satisfy the condition that
$\dd u/\dd z$ remains finite at the event horizon.
Selecting the solution with no radial nodes gives
$Q_\star\simeq0.8957024790$.
We denote the corresponding solution normalized to
$1$ at the event horizon by $u_\star$.
The relation $M_\star=(1+Q_\star^2)/2$ then gives
$M_\star\simeq0.9011414655$.

\subsection*{Step 2. Determine \texorpdfstring{$\gamma_\star$}{gamma star}.}

At $n=1$, $\Gamma_1$ and $\varkappa_1$ are determined by
$Q_\star$, $M_\star$, $g_{Q_\star}$, and $u_\star$
(see \cite[Appendix~C]{CompanionMath}).
We evaluate these coefficients numerically using the
results of Step~1.
Since $c_2=\Gamma_1+\varkappa_1\omega_4$, choosing
$\omega_4^\star=-\Gamma_1/\varkappa_1$ sets $c_2=0$.
For $Z_2$, \eqref{eq:omega-gamma-delta} therefore gives
\begin{equation*}
 \gamma_\star=1-\omega_4^\star
 =1+\frac{\Gamma_1}{\varkappa_1}
 \simeq1.9504423529.
\end{equation*}

\subsection*{Step 3. Determine \texorpdfstring{$\delta_\star$}{delta star}.}

With $\omega_4=\omega_4^\star$ and $c_2=0$, we carry out
the next recursion step to compute $\Gamma_2$ and
$\varkappa_2$, including the mass contribution
(see \cite[Appendix~C]{CompanionMath} for details).
The relation $c_4=\Gamma_2+\varkappa_2\omega_6$ then
gives $\omega_6^\star=-\Gamma_2/\varkappa_2$.
For $Z_3$, keeping $\gamma=\gamma_\star$ and choosing
\begin{equation*}
 \delta_\star=-1+2\gamma_\star-\omega_6^\star
 \simeq2.8556908431
\end{equation*}
sets $c_4=0$ without changing $c_2=0$.
The coefficients used in Steps~2 and~3 are listed in
Table~\ref{tab:tuning-coefficients};
the $n=2$ row is evaluated after setting
$\omega_4=\omega_4^\star$ so that $c_2=0$.

\begin{table}[t!] 
\centering
\caption{Coefficients at $\mu=1$.}
\label{tab:tuning-coefficients}
\setlength{\tabcolsep}{5pt}
\begin{tabular}{c r r r}
\hline\hline
$n$ & $\Gamma_n$ & $\varkappa_n$
    & $\omega_{2n+2}^{\star}$ \\
\hline
1 & $0.02279839533$ & $0.02398714163$ & $-0.9504423529$ \\
2 & $-0.001056104199$ & $0.02336831007$ & $0.04519386279$ \\
\hline\hline
\end{tabular} 
\end{table}

The resulting branches have contact orders $2$, $4$,
and $6$, respectively, as shown in
Fig.~\ref{fig:exponents}.

\section{Interior analysis}
We outline the continuation and estimates in every region from the event horizon to the
central singularity and the derivation of the Kasner asymptotics.
The technically involved complete nonlinear estimates are provided in
\cite[\S4]{CompanionMath}.
Throughout, $\mu>1/4$ is fixed, primes denote $\dd/\dd x$,
and $x_I=-2\ln Q_\star$ denotes the inner-horizon location
of the critical RN solution, used here as a reference.

\subsection*{Step 1. Continue from the event horizon.}

Let $\bar N_\star$ denote $\bar N$ for the RN solution
at $Q_\star$, with
$j=\bar N_\star\bar u_\star'$.
Fix $ L_+\in(0,x_I)$ independently of $\varepsilon$.
Horizon regularity and ordinary differential equation
continuation extend the solution to $[0,L_+)$, with
$\bar N,\bar J,\bar\psi>0$ for $x\in(0,L_+)$.
Smooth parameter dependence and parity give $\bar\psi(x;\varepsilon)
	 =\varepsilon\bar u_\star(x)+O(\varepsilon^3)$ and 
$\bar J(x;\varepsilon)
	 =\varepsilon j(x)+O(\varepsilon^3)$
together with $\bar N=\bar N_\star+O(\varepsilon^2)$ and
$\bar R=1+O(\varepsilon^2)$, uniformly on this fixed
interval as $\varepsilon\to0^+$.

\subsection*{Step 2. Reach the moving matching point.}

Set $x_m=x_I-\varepsilon$. The estimates in \cite[\S4.2]{CompanionMath} give 
$\bar N(x_m)\sim\lambda_I\varepsilon $,
	  $\bar J(x_m) \sim\iota\varepsilon $,  $
	\bar R(x_m)\sim 1  $ and  $\bar\psi(x_m)\lesssim \varepsilon|\ln\varepsilon|$ 
where $\lambda_I=1-Q_\star^2 
$, $ \iota=\lim_{x\to x_I^-}j(x)>0$. 

\subsection*{Step 3. Control the nonlinear process.}
We follow the solution until
$\bar R$ decreases (due to \eqref{eqA:interior}) to a fixed value $0<\eta<1$.
For sufficiently small $\varepsilon>0$, the first point
$x_\eta>x_m$ satisfying $\bar R(x_\eta)=\eta$ exists
and obeys $0<x_\eta-x_m\le4\varepsilon/\eta$ (see \cite[\S4.3]{CompanionMath}).
On $[x_m,x_\eta]$, $x$ remains close to $x_I$,
$Q(\varepsilon)$ is close to $Q_\star$, and
$\bar\psi$ is small. We can prove, approximately, $\mathcal A\approx -\lambda_I$ where $\mathcal A=\e^{-x}-Q(\varepsilon)^2\e^xW(\bar\psi)$. 
The equation for $\bar N$ in \eqref{eqA:interior}
therefore gives $\bar N'\approx-\lambda_I\bar R$.
Combining this with the equation for $\bar R'$ yields
$\dd\bar R/\dd\bar N\approx
\bar J^2/(\lambda_I\bar N^2)$.
Using $\bar J \sim \iota\varepsilon$ and integrating
this relation, we find that
\begin{equation}
 \mathcal I=\bar R+\frac{\bar J^2}{\lambda_I\bar N} \approx 1 \quad (\varepsilon\to0^+)
 \label{eq:transition-invariant}
\end{equation} 
for $x\in [x_m,x_\eta]$, which implies $\mathcal I(x_\eta)\approx \mathcal I(x_m)$. 
 
\begin{figure}[!htbp]
 \centering
 \includegraphics[width=\textwidth]{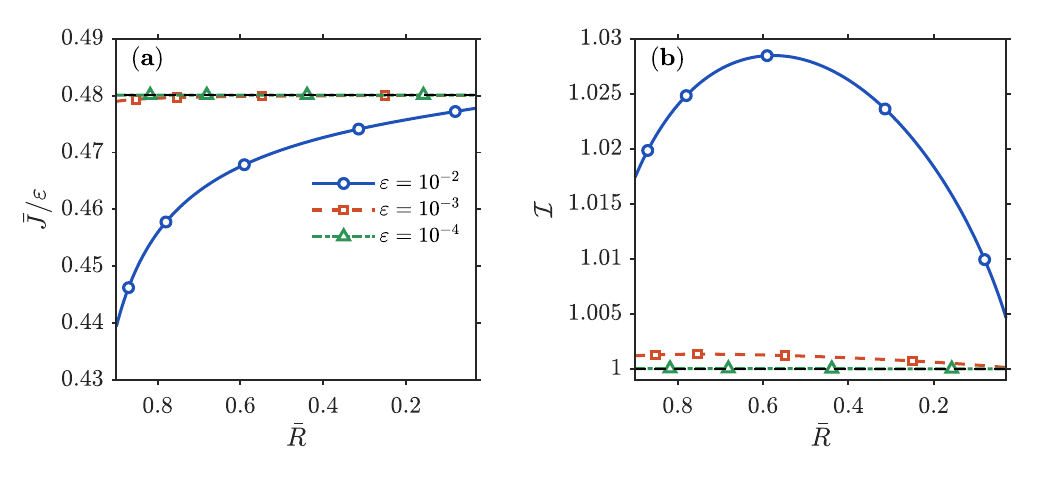}
\caption{For $\varepsilon=10^{-2},10^{-3},10^{-4}$, $Z(\psi)=1+\psi^2$ ($\mu=1$), the plots show $\bar J/\varepsilon$ and $\mathcal I$ in the interior region $0.03<\bar R<0.90$.\\
(a) As $\varepsilon$ decreases,
$\bar J/\varepsilon$ approaches $\iota$ (black dashed line).\\
(b) As $\varepsilon$ decreases, the quantity $\mathcal I=\bar R+\bar J^2/(\lambda_I\bar N)$ approaches 1 (black dashed line).\\
Since $\bar N=\bar J^2/[\lambda_I(\mathcal I-\bar R)]$,
these profiles illustrate how $\bar N$ reaches the
$O(\varepsilon^2)$ scale once $\mathcal I-\bar R$
has a positive lower bound independent of $\varepsilon$.}
\label{fig:transition}
\end{figure}

Since $\mathcal I(x_m)\to1$, we have
$\mathcal I(x_\eta)\to1$.
Using $\bar R(x_\eta)=\eta$ in
\eqref{eq:transition-invariant}, we therefore obtain $
 \frac{\bar N(x_\eta)}{\varepsilon^2}
 \rightarrow\frac{\iota^2}{\lambda_I(1-\eta)}$.  
Thus \textit{$\bar N$ decreases from order $\varepsilon$ at $x_m$
to order $\varepsilon^2$ at $x_\eta$}, whereas
$\bar J(x_\eta)\sim\iota\varepsilon$.
Consequently, $\bar\psi'=\bar J/\bar N$ is of order
$\varepsilon^{-1}$ at $x_\eta$.
Fig.~\ref{fig:transition} illustrates
$\bar J/\varepsilon$ and $\mathcal I$ for the couplings
$Z(\psi)=1+\psi^2$.
 
\subsection*{Step 4. Continue to the center.}
For any sufficiently small $\eta>0$, the solution
extends to $x\in[x_\eta,\infty)$ with $\bar N>0$.
For each fixed such $\varepsilon$, $\bar N$ and $\bar J$
approach finite positive limits $\bar N_c$ and $\bar J_c$
as $x\to\infty$.
The estimates in \cite{CompanionMath} give
$|\bar N_c-\bar N(x_\eta)|\le C\eta\varepsilon^2$
and
$0\le\bar J_c-\bar J(x_\eta)\le C\eta\varepsilon^2$.
Combining these bounds with the limits at $x_\eta$,
and taking $\varepsilon\to0^+$ at  small enough $\eta$,   gives $\bar N_c/\varepsilon^2\to
	\iota^2/\lambda_I$, $	\bar J_c/ \varepsilon\to\iota $.  
Since $\beta_{\rm K}=\bar J_c/\bar N_c$, we obtain \eqref{eq:intro_amplitude_scalings}.

\subsection*{Step 5. Recover the Kasner geometry.}

For each fixed sufficiently small $\varepsilon>0$,
the asymptotic analysis in \cite[\S4.4.2]{CompanionMath} gives
$\bar\psi=\beta_{\rm K}x+C_\psi+o(1)$ and
$\bar\chi=\beta_{\rm K}^2x+\chi_c+o(1)$
as $x\to\infty$.
The remaining proper time satisfies
$\dd\tau=-\sqrt{\bar R/\bar N}\e^{-3x/2}\dd x$.
Integrating this relation gives
$\tau\sim C_\tau\e^{-(\beta_{\rm K}^2+3)x/2}$.
Substitution into the metric \eqref{eq:ansatz} yields the Kasner form
and exponents in
\eqref{eq:kasner-metric}--\eqref{eq:kasner-exponents}.

\begin{figure}[!htbp]
 \centering
 \includegraphics[width=\textwidth]{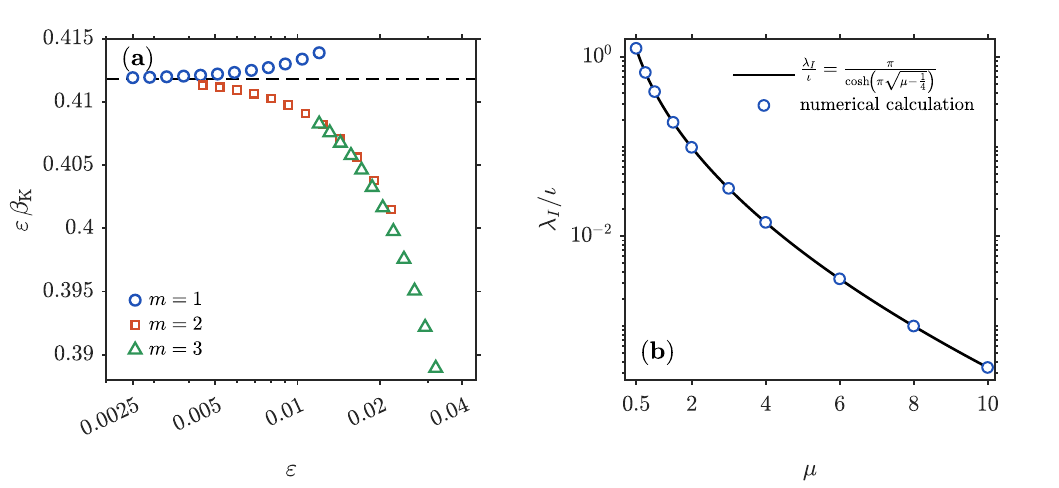}
\caption{Common inverse-amplitude interior response and its dependence on
the linear coupling.\\
(a) At $\mu=1$, $\varepsilon\beta_{\rm K}$ is plotted with $m=1,2,3$. The dashed line marks the predicted limit
$\lambda_I/\iota
=\pi/\cosh\!\left(\pi\sqrt{\tfrac34}\right)$.\\
(b) For $Z(\psi)=1+\mu\psi^2$, the points show numerical
estimates of $\lim_{\varepsilon\to0}\varepsilon\beta_{\rm K}$ for different $\mu$.  The solid black
curve is the exact relation
$\lambda_I/\iota
=\pi/\cosh\!\left(\pi\sqrt{\mu-\tfrac14}\right)$.}
\label{fig:inverse}
\end{figure}

At fixed $\mu$, the coefficient $\lambda_I/\iota$
is independent of the higher terms in $Z$.
Fig.~\ref{fig:inverse} (a) shows
$\varepsilon\beta_{\rm K}$ approaching the common limit
$\lambda_I/\iota$ for these couplings at $\mu=1$.
Fig.~\ref{fig:inverse} (b) shows the dependence of this
limit on $\mu$.
Complete proofs of global continuation and the absence
of an inner horizon are given in
\cite[\S~4]{CompanionMath}.

\section{Conclusions and discussions}
In this work, we  have identified the mechanism that selects the Kasner critical
exponent in the asymptotically flat scalarized black holes
considered here.
At fixed quadratic coupling parameter $\mu$, varying only the
higher-order terms in the scalar--electromagnetic coupling leaves
the scalarization threshold  $q_\star$  and the leading interior scaling law $\beta_{\rm K}\sim(\lambda_I/\iota)\varepsilon^{-1}$
unchanged. These variations can, however, change the leading power of
$\varepsilon$ in the exterior relation
$\Delta q=q(\varepsilon)/q_\star-1  \sim c_{2m}\varepsilon^{2m}$.  
For a branch with contact order $2m$, combining the interior scaling law and the exterior relation gives \eqref{eq:intro_critical_scaling}.

Thus the critical exponent measured in $\Delta q$ is set by
the first nonzero order in the exterior branch expansion.
The nondegenerate quadratic case $m=1$ analytically recovers
the $1/2$ law reported in \cite{LiSunYang2025CriticalInterior},
whereas successive fine-tuning of higher-order coupling
coefficients yields exponents $1/(2m)$ for any prescribed
integer $m\ge1$, including $1/4$ and $1/6$.

This scaling law also determines how the terminal Kasner
exponents approach their limiting values. 
Combining  \eqref{eq:intro_critical_scaling}
and \eqref{eq:kasner-exponents} gives $p_\Omega
	\sim
	2(\frac{\iota}{\lambda_I})^2
	|c_{2m}|^{-1/m}|\Delta q|^{1/m}$ and $1-p_t = 2p_\Omega $. 
Both $p_\Omega$ and $1-p_t$ therefore vanish as
$|\Delta q|^{1/m}$ when $q\to q_\star$ along the branch.
By contrast, the coefficient of the interior inverse-amplitude
law depends only on $\mu$
\begin{align}
	\frac{\lambda_I}{\iota}
	=
	\frac{\pi}
	{\cosh\!\left(\pi\sqrt{\mu-\tfrac14}\right)},
	\qquad \mu>\frac14.
	\label{eq:closed-form-amplifier}
\end{align}
For $\mu=1$, this gives
$\lambda_I/\iota\simeq0.4118294226$, consistent with the
small-amplitude behavior in Fig.~\ref{fig:inverse}.
Thus models with the same scalarization threshold and the
same leading interior amplitude law can nevertheless exhibit
different critical exponents.
The exponent contains nonlinear information about the exterior
branch that cannot be inferred from the scalarization
threshold alone.

Since this Letter is a brief report focused on the main results,
we present the complete but very lengthy nonlinear mathematical analysis in a
separate companion work \cite{CompanionMath}.
That work will include the derivation of \eqref{eq:closed-form-amplifier} and detailed proofs of
the scalarized branch construction, the higher-order contact
recursion, the global interior continuation, and the Kasner
asymptotics. 

\appendix

\section{Radial equations}\label{app:radial}
The complete derivations and proofs  in Appendixes \ref{app:radial}--\ref{app:tuning} are technically involved and require substantial
mathematical machinery. Here we present only the outlines and main steps. Full
details are given in a long mathematical article \cite{CompanionMath}. 

Varying the action in \eqref{eq:action} and substituting
the metric in \eqref{eq:ansatz}, we obtain the following
exterior equations, with $g=f/z^2$ and $W=Z^{-1}$
\begin{align}
 \frac{\dd\chi}{\dd z}
 &=z\left(\frac{\dd\psi}{\dd z}\right)^2,\label{dchi}\\
 \frac{\dd g}{\dd z}
 &=\frac{g-1}{z}
   +zg\left(\frac{\dd\psi}{\dd z}\right)^2
   +Q^2zW,\\
 \frac{\dd}{\dd z}
 \left(\e^{-\chi}g\frac{\dd\psi}{\dd z}\right)
 &=\tfrac12 Q^2\e^{-\chi}W_{,\psi},
 \label{eqA:exterior}
\end{align}
where $Q$ and $M$ are the electric charge and ADM mass
in event-horizon-radius units.
For fixed $Z$ and $\varepsilon$, we impose
$g(1)=0$, $\chi(1)=0$, and $\psi(1)=\varepsilon$,  
and  require the field to remain finite and smooth at the event horizon.
 We determine $Q$ by requiring the resulting solution to satisfy
$\psi(0)=0$ at spatial infinity (see the shooting method in \cite[\S3]{CompanionMath}).   
The corresponding mass $M$ is obtained from
$g(z)=1-2Mz+O(z^2)$ as $z\to0$.
With primes denoting $\dd/\dd x$, the interior equations are
\begin{equation}
 \begin{aligned}
  \bar\psi'&=\bar J/\bar N,
  &\bar R'&=-(\bar J/\bar N)^2\bar R,\\
  \bar N'&=\bar R(\e^{-x}-Q^2\e^xW),
  &\bar J'&=-\tfrac12Q^2\e^x\bar R W_{,\psi}.
 \end{aligned}
 \label{eqA:interior}
\end{equation}

\section{Exterior analysis and derivation of the recursive relation}\label{app:tuning}
We derive the recursive relation between the higher-order
coupling coefficients and the exterior expansion coefficients.
In particular, we show that
$c_{2n}=\Gamma_n+\varkappa_n\omega_{2n+2}$ with
$\varkappa_n>0$.
The detailed calculations are given in
\cite[Appendix~C]{CompanionMath}, and we only outline the
main steps here.

\subsection*{Step 1. Expand in \texorpdfstring{$s$}{s}.}
Fix an integer $m\ge1$ and $\mu>1/4$, and use the coupling
expansion \eqref{eq:coupling-expansion}.
Since $\psi|_{\varepsilon=0}=0$, Taylor's theorem allows us to write
$\psi=\varepsilon U$.
At $\varepsilon=0$, $g,\chi,Q$, and $M$ take their critical
RN values, while the leading-order scalar equation
\eqref{eq:zero-mode} gives $U|_{\varepsilon=0}=u_\star$.
The branch regularity and parity then yield the following
finite expansions in $s=\varepsilon^2$ as $s\to0^+$
\begin{equation} 
  U(z;s)=u_\star(z)+\sum_{n=1}^{m}u_{2n+1}(z)s^n
          +O(s^{m+1}).
 \label{eqC:branch-expansions}
\end{equation}
The quantities $g,\chi,Q$, and $M$ have analogous expansions
to the same order, with respective coefficients
$g_{2n}(z)$, $\chi_{2n}(z)$, $Q_{2n}$ and $M_{2n}$, and zeroth-order
terms $g_{Q_\star}(z)$, $0$, $Q_{\star}$ and $M_\star$ from the critical
RN solution.
Substitution into \eqref{dchi}--\eqref{eqA:exterior}
gives equations at successive orders in $s$.

\subsection*{Step 2. Determine the metric coefficients.}

Suppose that all branch coefficients through order $s^{n-1}$
are known.
Substituting \eqref{eqC:branch-expansions} and
$\psi=\varepsilon U$ into the metric equations and comparing
coefficients of $s^n$ gives the equations for
$\chi_{2n}$ and $g_{2n}$.
The equation for $\chi_{2n}$ contains only known source terms;
integrating it with $\chi_{2n}(1)=0$ determines $\chi_{2n}$.
The equation for $g_{2n}$ still contains the new charge
coefficient $Q_{2n}$.
Integrating it with $g_{2n}(1)=0$ gives
\begin{align}
g_{2n}(z)
=
\widetilde g_{2n}(z)
+2Q_\star Q_{2n}z(z-1),
\label{eqC:metric-split}
\end{align}
where $\widetilde g_{2n}$ depends only on lower-order coupling
and branch coefficients.
Thus $g_{2n}$ is determined apart from $Q_{2n}$.
The new coupling coefficient $\omega_{2n+2}$ enters the
metric equation directly only at order $s^{n+1}$. 

\subsection*{Step 3. Determine \texorpdfstring{$Q_{2n}$}{Q(2n)}.}

Substituting $\psi=\varepsilon U$ into
\eqref{eqA:exterior} and dividing by $\varepsilon$ gives
an equation for $U$.
We insert the coupling and branch expansions and equate
the coefficients of $s^n$ on both sides.
Using the metric decomposition \eqref{eqC:metric-split}
and collecting the terms involving the new unknowns
$u_{2n+1}$ and $Q_{2n}$ yields
\begin{equation} 
  \frac{\dd}{\dd z}
  \left[g_{Q_\star}(z)\frac{\dd u_{2n+1}}{\dd z}\right]
  +\mu Q_\star^2u_{2n+1} 
  =F_{\psi,2n+1}-Q_{2n}F_Q.
 \label{eqC:scalar-recursion}
\end{equation}
Here $F_Q$ is a known function, and $F_{\psi,2n+1}$
is determined by lower-order data together with
$\omega_{2n+2}$.
For a given $\omega_{2n+2}$, the Fredholm solvability
condition uniquely determines $Q_{2n}$.

\subsection*{Step 4. Obtain \texorpdfstring{$u_{2n+1}$}{u(2n+1)} and \texorpdfstring{$M_{2n}$}{M(2n)}.}

With $Q_{2n}$ fixed, we solve
\eqref{eqC:scalar-recursion} to derive $u_{2n+1}$.
The mass coefficient follows from
$g(z)=1-2Mz+O(z^2)$ at spatial infinity. Then the coefficient
of $z$ in the expansion of $g_{2n}(z)$ is $-2M_{2n}$.
The term $2Q_\star Q_{2n}z(z-1)$ in
\eqref{eqC:metric-split} therefore contributes
$Q_\star Q_{2n}$ to the mass coefficient, giving
\begin{align}
M_{2n}=\widetilde M_{2n}+Q_\star Q_{2n}.
\label{eqC:mass-split}
\end{align}
Here $\widetilde M_{2n}$ is read from $\widetilde g_{2n}$
in the same way and depends only on lower-order data,
so it is independent of $\omega_{2n+2}$.
This completes the recursion at order $s^n$.

\subsection*{Step 5. Determine \texorpdfstring{$c_{2n}$}{c(2n)}.}

Expanding $(M/M_\star)(q/q_\star)=Q/Q_\star$, comparing the
coefficients of $s^n$ and considering $Q_{2n}$ from Step $3$ and $M_{2n}$ from \eqref{eqC:mass-split} give
\begin{equation*}
 c_{2n}=\frac{Q_{2n}}{Q_\star}
       -\frac{M_{2n}}{M_\star}
       -\sum_{k=1}^{n-1}\frac{M_{2k}}{M_\star}c_{2(n-k)}
       \quad\Rightarrow\quad\eqref{eq:triangular}.
\end{equation*} 
Varying $\omega_{2n+2}$ therefore changes $c_{2n}$
with nonzero slope without changing
$c_2,\ldots,c_{2n-2}$.

\section*{Acknowledgments}
This work is partially supported by NSFC (Grant No.
12571234) and Fundamental Research Funds for the Central Universities of HUST (Grant No. 2025BRSXA001). 

\section*{Data availability}
There are no publicly available
research data or software supporting this manuscript.
Requests for further information or data should be sent
to the authors.

\bibliographystyle{amsplain}
\bibliography{references}

\providecommand{\bysame}{\leavevmode\hbox to3em{\hrulefill}\thinspace}
\providecommand{\MR}{\relax\ifhmode\unskip\space\fi MR }
\providecommand{\MRhref}[2]{%
  \href{http://www.ams.org/mathscinet-getitem?mr=#1}{#2}
}
\providecommand{\href}[2]{#2}
\begin{thebibliography}{10}

\bibitem{AstefaneseiEtAl2019EMSClasses}
Dumitru Astefanesei, Carlos Herdeiro, Alexandre Pombo, and Eugen Radu,
  \emph{{Einstein--Maxwell--scalar} black holes: classes of solutions, dyons
  and extremality}, Journal of High Energy Physics \textbf{10} (2019), 078.

\bibitem{BelkhadriaPombo2024MixedScalarization}
Zakaria Belkhadria and Alexandre~M. Pombo, \emph{Mixed scalarization of charged
  black holes: From spontaneous to nonlinear scalarization}, Physical Review D
  \textbf{110} (2024), 044014.

\bibitem{CaiLiYang2021NoInnerHorizon}
Rong-Gen Cai, Li~Li, and Run-Qiu Yang, \emph{No inner-horizon theorem for black
  holes with charged scalar hairs}, Journal of High Energy Physics \textbf{03}
  (2021), 263.

\bibitem{DiasHorowitzSantos2021InsideAF}
Oscar J.~C. Dias, Gary~T. Horowitz, and Jorge~E. Santos, \emph{Inside an
  asymptotically flat hairy black hole}, Journal of High Energy Physics
  \textbf{12} (2021), 179.

\bibitem{DonevaEtAl2024ScalarizationReview}
Daniela~D. Doneva, Fethi~M. Ramazano\u{g}lu, Hector~O. Silva, Thomas~P.
  Sotiriou, and Stoytcho~S. Yazadjiev, \emph{Spontaneous scalarization},
  Reviews of Modern Physics \textbf{96} (2024), 015004.

\bibitem{FernandesEtAl2019CouplingDependence}
Pedro G.~S. Fernandes, Carlos A.~R. Herdeiro, Alexandre~M. Pombo, Eugen Radu,
  and Nicolas Sanchis-Gual, \emph{Spontaneous scalarisation of charged black
  holes: Coupling dependence and dynamical features}, Classical and Quantum
  Gravity \textbf{36} (2019), 134002, Erratum: Class. Quantum Grav. 37 (2020)
  049501.

\bibitem{HartnollEtAl2021Diving}
Sean~A. Hartnoll, Gary~T. Horowitz, Jorrit Kruthoff, and Jorge~E. Santos,
  \emph{Diving into a holographic superconductor}, SciPost Physics \textbf{10}
  (2021), 009.

\bibitem{CompanionMath}
Chihang He and Chao Liu, \emph{Proofs on {Kasner} critical scaling in
  {Einstein-Maxwell-scalar} black holes with nonlinear collapse and
  scalarization}, In preparation.

\bibitem{Henneaux2022FinalKasner}
Marc Henneaux, \emph{The final {Kasner} regime inside black holes with scalar
  or vector hair}, Journal of High Energy Physics \textbf{03} (2022), 062.

\bibitem{HerdeiroEtAl2018ChargedScalarization}
Carlos A.~R. Herdeiro, Eugen Radu, Nicolas Sanchis-Gual, and Jose~A. Font,
  \emph{Spontaneous scalarization of charged black holes}, Physical Review
  Letters \textbf{121} (2018), 101102.

\bibitem{Hod2020ExistenceLine}
Shahar Hod, \emph{Spontaneous scalarization of charged {Reissner--Nordstr\"om}
  black holes: Analytic treatment along the existence line}, Physics Letters B
  \textbf{798} (2019), 135025.

\bibitem{LiSunYang2025CriticalInterior}
Li~Li, Ze~Sun, and Fu-Guo Yang, \emph{Critical phenomenon inside asymptotically
  flat black holes with spontaneous scalarization}, Physical Review Letters
  \textbf{136} (2026), 251402.

\bibitem{Ori1991MassInflation}
Amos Ori, \emph{Inner structure of a charged black hole: an exact
  mass-inflation solution}, Physical Review Letters \textbf{67} (1991),
  789--792.

\bibitem{PoissonIsrael1990InternalStructure}
Eric Poisson and Werner Israel, \emph{Internal structure of black holes},
  Physical Review D \textbf{41} (1990), 1796--1809.

\bibitem{VanDeMoortel2024ViolentCollapse}
Maxime Van~de Moortel, \emph{Violent nonlinear collapse in the interior of
  charged hairy black holes}, Archive for Rational Mechanics and Analysis
  \textbf{248} (2024), 89.

\end{thebibliography}

\end{document}